\documentclass[12pt]{iopart}

\usepackage{iopams}
\usepackage{bm,bbm}
\usepackage{amsfonts}
\usepackage{amssymb}
\usepackage{amscd}
\usepackage{dsfont}
\usepackage{cite}
\usepackage[dvips]{graphicx}
\usepackage{euscript}
\usepackage{pstricks}

\newcommand{\set}[1]{\left\{#1\right\}}
\newcommand{\dm}[1]{d\mu\left(#1\right)}
\newcommand{\eps}{\varepsilon}

\renewcommand{\tr}{{\rm Tr}}

\begin{document}

\title[Random Bures mixed states and the distribution of their purity]{Random Bures mixed states \\
           and the distribution of their purity}

\author{V.A. Osipov$^1$, H.-J. Sommers$^1$, and K. {\.Z}yczkowski$^{2,3}$}.

\address{\it $^1$Fachbereich Physik, Universit\"{a}t Duisburg-Essen \\
47048 Duisburg, Germany}
\address{$^2$Institute of Physics,   Jagiellonian University, \\
 ul. Reymonta 4, 30-059 Krak{\'o}w, Poland}
\address{$^3$Center for Theoretical Physics, Polish Academy of Sciences\\
         Al. Lotnik{\'o}w 32/44, 02-668 Warszawa, Poland}
\ead{Vladimir.Al.Osipov@gmail.com \quad H.J.Sommers@uni-due.de \quad karol@cft.edu.pl }

\date{September 14, 2009}

\begin{abstract}
Ensembles of random density matrices determined by
various probability measures are analysed.
A simple and efficient algorithm to generate
at random density matrices distributed according to the Bures measure is proposed. This procedure may serve as an initial step in performing
Bayesian approach to quantum state estimation based on the Bures prior.
We study the distribution of purity of random mixed states.
The moments of the distribution of purity are determined
for quantum states generated with respect to the Bures measure.
This calculation serves as an exemplary application of the ``deform-and-study'' approach based on ideas of integrability
theory. It is shown that Painlev\'e equation appeared as a part of the presented theory.
\end{abstract}

\submitto{\JPA}

\section{Introduction}

Random density matrices are a subject of a large current interest.
In some cases one considers ensembles of random pure states
defined on a finite dimensional Hilbert space ${\cal H}_K$.
A natural ensemble is defined by the Fubini-Study measure $\mu_{FS}$,
which is induced by the Haar measure on the unitary group $U(K)$
and invariant with respect to unitary rotations.

In some cases one needs to consider ensembles of
mixed quantum states. If the dimensionality $K$ is a composite number
$K=MN$ then an ensemble of random mixed
states can be obtained by partial trace over an $M$-dimensional subsystem,
$\rho={\rm Tr}_M|\psi\rangle \langle \psi|$.
If random pure states $|\psi\rangle$ are distributed according to $\mu_{FS}$,
one obtains in the set of density matrices of order $N$
a family of {\sl induced measures} \cite{Br96,Ha98,ZS01}, denoted here by $\mu_{N,M}$
In the symmetric case, $M=N$, the induced measure
is equal to the {\sl Hilbert-Schmidt measure},
which covers the entire set $\Omega$ of the density matrices
and is determined by the HS metric.

This observation leads to a simple algorithm to generate a
Hilbert-Schmidt random matrix \cite{ZS01}:
a) Take a square complex random matrix $A$ of size $N$ pertaining to the
Ginibre ensemble \cite{Gi,Me91} (with real and imaginary parts of
each element being independent normal random variables);
b) Write down the random matrix
\begin{equation}
\rho_{\rm HS} \ = \  \frac{\ \ AA^{\dagger}} {{\rm Tr}AA^\dagger} ,
\label{hsrand}
\end{equation}
which is by construction Hermitian, positive definite
and normalised, so it forms a legitimate density matrix.
Observe that the Ginibre matrix $A$ can be used to
represent a  random pure state of a bipartite system
 in a product basis,
$|\psi\rangle=\sum_{i,j} A_{ij}|i\rangle \otimes |j\rangle$.
The above procedure is thus equivalent to setting
\begin{equation}
\rho_{\rm HS} \ = \  {\rm Tr}_N |\psi\rangle \langle \psi | \; ,
\quad {\rm where \quad}
|\psi\rangle \in {\cal H}_N \otimes {\cal H}_N
\label{hsrand2}
\end{equation}
is a normalized random state taken from the composite Hilbert space
of size $N^2$ according to the Fubini--Study measure,
while Tr$_N$ denotes the partial trace over the second
$N$--dimensional subsystem.

Another distinguished measure in the space $\Omega$ of quantum mixed states,
is induced by the Bures metric \cite{Bu69,Uh92},
\begin{equation}
D_B(\rho,\sigma)=
\sqrt{ 2- 2{\rm Tr}(\sqrt{\rho}\sigma \sqrt{\rho})^{1/2}} .
\label{bures1}
\end{equation}
This metric induces the Bures probability distribution,
defined by the conditions that any ball with respect to the Bures distance
of a fixed radius in the space of quantum states has the same measure.
The Bures metric, related to quantum distinguishability \cite{Fu96},
plays a key role in analyzing the space of quantum states \cite{BZ06}.
The Bures metric is known to be the minimal monotone metric \cite{PS96}
and applied to any two diagonal matrices it gives their statistical distance.
These unique features of the Bures distance support the claim
that without any prior knowledge on a certain density matrix
acting on space ${\cal H}_N$,
the optimal way to mimic it is to generate the state at
random with respect to the Bures measure.

More formally, trying to reconstruct the quantum state out of
the results of the measurement \cite{He76,BDDAW,HRFJ04},\cite[chapt. 3]{Hy06}
one can follow the Bayesian mean estimation \cite{SBC01,CFS02}.
In this approach one starts selecting
a {\sl prior} probability distribution $P_0$ over the set $\Omega$
of all quantum states.
Acquiring experimental data one
uses them to generate likelihood function,
multiplies it by the prior and  normalizes the result
to obtain a posterior probability distribution $P_1$.
This distribution reflects the knowledge of an estimator,
so the best estimation of the quantum state
is given by the mean state with respect to
this distribution, $\rho_1=\int_{\Omega}  \rho P_1(\rho) d\rho$.
If more experimental data are gathered
one continues with this procedure
to obtain further probability distributions $P_n(\rho)$
and a sequence of expected states,
$\rho_n=\int_{\Omega} \rho P_n(\rho) d\rho$,
with $n=2,3, \dots$. This iterative procedure
should yield an accurate estimate of the unknown state   \cite{BK06}.

As the starting point    for such a reconstruction procedure
one  should chose as uninformative ("uniform")
distribution $P_0$ as possible,
so  the {\sl Bures prior} is often used for this purpose
\cite{Sl96,BMTR04,BBMTR05}.
In practice Bayesian method relays on computing integrals
over the set $\Omega$ of quantum states. Since analytical integration
is rarely possible, one needs to apply  some variants of the
numerical Monte Carlo method.  For this purpose an efficient algorithm
of generating random states according to a given distribution is required.
Although the Bures measure was investigated in several recent
papers \cite{Ha98,Sl99b,Sl01,SZ03,SZ04,ZS05},
no simple method to generate states with respect to this
measure was known.

Main aim of this work is to solve a few open problems related
to the Bures measure. We construct the following algorithm
to generate random Bures states:
a) Take a complex random matrix $A$ of size $N$ pertaining to the
 Ginibre ensemble and a random unitary matrix $U$
  distributed according to the Haar measure on $U(N)$ \cite{PZK98,Me07}.
b) Write down the random matrix
\begin{equation}
\rho_{\rm B} \ = \  \frac{\ \ ({\mathbbm 1}+U)AA^{\dagger} ({\mathbbm 1}+U^{\dagger})}
{{\rm Tr}[({\mathbbm 1}+U)AA^{\dagger} ({\mathbbm 1}+U^{\dagger})]} \,
\label{burrand}
\end{equation}
which is proved to represent a normalized quantum state
distributed according to the Bures measure.
In analogy to the Hilbert-Schmidt case we may also write
\begin{equation}
\rho_{\rm B} \ = \ \frac{{\rm Tr}_N |\phi\rangle \langle \phi |}
{\langle \phi|\phi\rangle }
\quad {\rm where \quad}
|\phi\rangle := [({\mathbbm 1}+U)\otimes {\mathbbm 1}]|\psi\rangle \; ,
\label{burrand2}
\end{equation}
$U\in U(N)$ and $|\psi\rangle$ is a random state of a bipartite system
used in  eq. (\ref{hsrand2}).
A similar construction is also provided to obtain random real Bures matrices.

The degree of mixture of any state $\rho$ of size $N$
 can  easily be  characterised by its purity $P(\rho)={\rm Tr}\rho^2$.
This quantity varies from  $1/N$ for the maximally mixed state, ${\mathbbm 1}/N$,
to unity, characteristic of an arbitrary pure state.
Characterisation of purity of random states,
related to the entanglement of initially pure states before the reduction,
is a subject of a considerable current interest \cite{Gi07,Zn07,Gi07c}.
The average purity is known for random states distributed with respect to
induced measures, \cite{Lu78,ZS01}, and for the Bures measure \cite{SZ04}
but the distribution of purity is known
only for the HS measure for low dimensions \cite{Gi07c}.
For the induced measures the moments of purity
were obtained in a recent work of Giraud \cite{Gi07}.
These results can be rederived by a method involving the methods of theory of
integrable systems (see~\cite{AvM} and also explanations in part~\ref{part6} of this article),
 which allows to obtain a recurrence relation between moments by deriving a differential equation
for the corresponding generation function.
 This differential equation is the IV-th Painlev\'e transcendent~\cite{FW}.
Since these moments are already known in the literature
we will concentrate on a more involved case
and derive the moments of the purity with respect to Bures measure.
Our calculations demonstrate practical usefulness of
this analytic technique and suggest, it might also be
used in solving related problems.

This paper is organised as follows.
In section~\ref{Sec2} we review probability measures in the space of mixed
quantum states and provide necessary definitions.
In section~\ref{Sec3} we derive a useful representation of the Bures measure
which allows us to construct the algorithm based on eq.(\ref{burrand}).
A similar reasoning is provided in section~\ref{Sec4} for real density matrices.
In section~\ref{Sec5} we analyse the moments of purity for a general class of
probability measures.
Results obtained there are used in section~\ref{Sec6} to derive explicit results on the moments
of purity for Bures random states. Some auxiliary calculations are relegated to the appendix.

\section{Ensembles of random density matrices}\label{Sec2}

We are going to analyse ensembles of random states, for which the probability
measure has a product form and may be factorised \cite{Ha98,ZS01},
\begin{equation}
  {\rm d} \mu_{\rm x} \ =\  {\rm d} \nu_{\rm x}
 (\lambda_1,\lambda_2,...,\lambda_N) \times {\rm d} \mu_U,
  \label{product}
\end{equation}
so the distribution of eigenvalues and eigenvectors are independent.
It is natural to assume that the eigenvectors
are distributed according to the unique, unitarily invariant,
Haar measure  $d\mu_U$  on $U(N)$. Taking this assumption as granted
the measure in the space of density matrices will be determined
by the first factor ${\rm d} \nu_{\rm x}$  describing the
distribution of eigenvalues $P(\lambda)$.

Consider a class of induced measures $\mu_{N,M}$
in the space of density matrices of size $N$.
To generate a mixed state according to such a measure one may
take a random bipartite $N\times M$  pure state $|\psi\rangle$,
(e.g. an eigenstate of a random Hamiltonian),
and trace out the $M$-dimensional environment.
This procedure yields the following probability distribution
\begin{equation}
\label{dmr}
    d\mu (\rho) \ \propto \
\Theta (\rho) \ \delta(\Tr \rho -1) \det \rho^{M-N}.
\end{equation}
It reflects the properties of density matrices $\rho \ge 0$ and $ \Tr \rho =1$.
In the special case $M=N$  the term with the determinant is equal to unity
and the induced measure reduces to the Hilbert-Schmidt measure.
The matrix $\rho$ is Hermitian and integrating out the eigenvectors of $\rho$ one reduces
$d\mu$ to the measure on the simplex of eigenvalues $\{ \lambda_1,...\lambda_N\}$ of
the density matrix \cite{Lu78},
\begin{equation}
\label{dmuM}
d\mu_M(\lambda_1,...\lambda_N)  = C_{N,M} \;
 \delta\left( \sum_i\lambda_i -1\right)
\Delta^2_N(\bm \lambda)
  \prod_i \Theta(\lambda_i) \lambda_i^{M-N} {\rm d}\lambda_i
\end{equation}
where the squared Vandermonde determinant
\begin{equation}
\label{vdm}
\Delta_N^2(\bm \lambda) \; := \;
 \prod_{i<j}^{1...N} (\lambda_i -\lambda_j )^2
\end{equation}
  appears as a geometric consequence of diagonalisation. The normalisation constant
\begin{equation}
C_{N,M} = \frac{\Gamma(MN)} {\prod_{j=0}^{N-1} \Gamma(M-j) \Gamma(N-j+1) }
\label{cnm}
\end{equation}
 has been calculated in \cite{ZS01}.

Furthermore, we analyse the measure induced by the Bures distance,  which
is characterised by the following probability of eigenvalues  \cite{Ha98}
\begin{equation}
\label{dmuB}
  d\mu_B(\lambda_1,...\lambda_N) =
  C_N^B \; \delta\left( \sum_i\lambda_i -1\right)
\prod_i \Theta(\lambda_i) \lambda_i^{-1/2} {\rm d}\lambda_i
\prod_{i<j}^{1...N}
\frac{(\lambda_i -\lambda_j )^2}{ \lambda_i +\lambda_j}
\end{equation}
The normalisation constant for this measure
\begin{equation}
C_N^B =
2^{N^2-N}\ \frac{\Gamma(N^2/2)}
  {\pi^{N/2}\  \prod_{j=1}^{N} \Gamma(j+1) }
\label{burcn}
\end{equation}
was obtained in \cite{Ha98,Sl99b} for small  $N$ and in \cite{SZ03}
in the general case. It is easy to see that the case $N=2$ is somewhat special since
the denominator in the last factor is equal to unity. Incidentally, the Bures measure coincides in this case
with the induced measure with an unphysical
half-integer dimension of the environment, $M=3/2$,
but this observation may ease some computations  \cite{ZS05}.

\section{Generating Bures density matrices}\label{Sec3}

In this section we show that eq. (\ref{burrand}) may be used to construct an
ensemble of random states distributed according to the Bures measure. To this end we will
rewrite the Bures probability distribution corresponding to the measure  (\ref{dmuB})
in a more suitable form, which involves random unitary matrices.

As a warm-up we shall first consider the induced measure.
Let us start with a probability measure defined by an integral
over random matrices $A$ with respect to the Ginibre measure,
 $\exp(-{\rm Tr} AA^{\dagger})$,
  \begin{equation}
\label{pmuhs1}
P_M(\rho) \propto \int dA e^{-{\rm Tr} AA^{\dagger}}
\ \delta \left(  \rho - \frac{AA^{\dagger}}{{\rm Tr} AA^{\dagger}} \right)  .
\end{equation}
Here $A$ denotes a rectangular complex matrix of dimension $N \times M$,
and it is assumed that $M \ge N$.
Let us introduce another $\delta$-function by an integral with respect to an auxiliary variable $s$,
\begin{equation}
\label{pmuhs2}
P_M(\rho) \propto \int_0^{\infty} d s  \int dA e^{-{\rm Tr} AA^{\dagger}}
\ \delta \left(  \rho - \frac{AA^{\dagger}}{s}\right)  \;
 \delta ( s - {\rm Tr} AA^{\dagger}).
\end{equation}
After rescaling the matrix variable, $A \to \sqrt{s} A$,
the above equation takes the form
\begin{eqnarray}
\label{pmuhs3}
P_M(\rho) & \; \propto \; \int_0^{\infty} \frac{d s}{s} e^{-s} s^{MN}
  \int dA \; \delta \left(  \rho - AA^{\dagger}\right)
 \delta \left(  1 - {\rm Tr} AA^{\dagger}\right)  \nonumber \\
& \;  \propto \; \Theta(\rho)\; ({\rm det} \rho)^{M-N} \;
 \delta ( 1 - {\rm Tr} \rho)
\end{eqnarray}
This form is equivalent to (\ref{dmr}), what
proves that random matrices distributed according to the
induced measure can be generated from rectangular complex matrices of the Ginibre ensemble.
Taking in particular square $N \times N$ matrices one generates
random Hilbert-Schmidt  states according to (\ref{hsrand}).

To repeat this reasoning for the Bures matrices we will start with
a similar ensemble defined by a double integral
\begin{equation}
\label{pbur1}
P_B(\rho) \propto \int dA \int dH e^{-{\rm Tr}[AA^{\dagger}+ H^2 AA^{\dagger}]}
\ \delta \left(  \rho - \frac{AA^{\dagger}}{{\rm Tr} AA^{\dagger}}\right)  .
\end{equation}
Here $A$ can be interpreted as a $N \times M$ Ginibre random matrix,
while $H$ is a Hermitian matrix of order $N$.
As in the earlier case we introduce a $\delta$-function by integrating over
an auxiliary variable $s$
\begin{equation}
\label{pbur2}
P_B(\rho) \propto \int_0^{\infty} d s \int dA \int dH
 e^{-{\rm Tr}[AA^{\dagger}+ H^2 AA^{\dagger}]}
\ \delta \left(  \rho - \frac{AA^{\dagger}}{s} \right)
 \delta ( s - {\rm Tr} AA^{\dagger}).
\end{equation}
Rescaling $A \to \sqrt{s} A$ leads to
\begin{equation}
\label{pbur3}
P_B(\rho) \propto \int_0^{\infty}
\frac{d s}{s} e^{-s} s^{MN}
 \int dH e^{- s {\rm Tr}H^2 \rho}
\int dA \delta ( \rho - AA^{\dagger})
\delta ( 1 - {\rm Tr} \rho) .
\end{equation}
Performing another rescaling,
$H \to H/\sqrt{s} $, we arrive at
\begin{equation}
\label{pbur4}\fl
P_B(\rho) \propto \int_0^{\infty}
\frac{d s}{s} e^{-s} s^{MN-N^2/2}
\prod_i  \frac{1}{\sqrt{\lambda_i}}
\prod_{i <j}   \frac{1}{\lambda_i+\lambda_j}
\Theta(\rho)  \delta ( 1 - {\rm Tr} \rho)
({\rm det} \rho)^{M-N} .
\end{equation}
Note that the integration over $s$ gives a constant factor only, which
will be absorbed into the proportionality relation, while
integration over eigenvectors of $\rho$ gives the squared Vandermonde determinant.
Furthermore, in the case $M=N$ the last factor equals to unity, so the above
expression reduces to the  Bures measure (\ref{dmuB}).

Let us then return to the starting integral (\ref{pbur1})
and apply another rescaling, $A \to \frac{1}{\sqrt{{\mathbbm 1}+H^2}}A$.
It leads to the following expression
\begin{equation}
\label{pbur5}
P_B(\rho) \propto \int \frac{d H}{[{\rm det}({\mathbbm 1}+H^2)]^M}
\int dA e^{-{\rm Tr}AA^{\dagger}}
\ \delta \left(  \rho - \frac{   \frac{{\mathbbm 1}}{{\mathbbm 1}+iH}  AA^{\dagger} \frac{{\mathbbm 1}}{{\mathbbm 1}-iH}}
{{\rm Tr} \frac{{\mathbbm 1}}{{\mathbbm 1}+iH}  AA^{\dagger} \frac{{\mathbbm 1}}{{\mathbbm 1}-iH}}
 \right) .
\end{equation}
At this point it is convenient to introduce
an unitary variable matrix
$$
U=\frac{{\mathbbm 1} - iH}{{\mathbbm 1} + iH}.
$$
As shown in lemma 1 proved in \ref{app1}
the 'Cauchy--like' measure
$d H / [{\rm det}({\mathbbm 1}+H^2)]^N$
is equivalent to the Haar measure $d\mu(U)$ on $U(N)$.
Moreover, since
$$
\frac{{\mathbbm 1}}{{\mathbbm 1}+iH}=\frac{{\mathbbm 1}+U}{2},
$$
the above expression is equivalent to
\begin{equation}
\label{pbur6}
P^R_B(\rho) \propto \int_{U(N)} d\mu(U)
\int dA e^{-{\rm Tr}AA^{\dagger}}
\ \delta \left(  \rho - \frac{   \frac{{\mathbbm 1}+U}{2}  AA^{\dagger} \frac{{\mathbbm 1}+U^{\dagger}}{2}}
{{\rm Tr}   \frac{{\mathbbm 1}+U}{2}
AA^{\dagger} \frac{{\mathbbm 1}+U^{\dagger}}{2}}
 \right)  .
\end{equation}
The factors $1/2$ cancel out, so taking a square complex random Ginibre matrix $A$
and a random unitary matrix $U$ of the same size we can generate random Bures
matrices according to the constructive recipe (\ref{burrand}).
Writing a random state $|\psi\rangle$  in a product basis,
$|\psi\rangle=\sum_{i,j} A_{ij}|i\rangle \otimes |j\rangle$,
we infer that this method of generating random Bures
states may alternatively be written by eq. (\ref{burrand2}).

\begin{figure} [htbp]
   \begin{center}
 \includegraphics[width=9.0cm,angle=0]{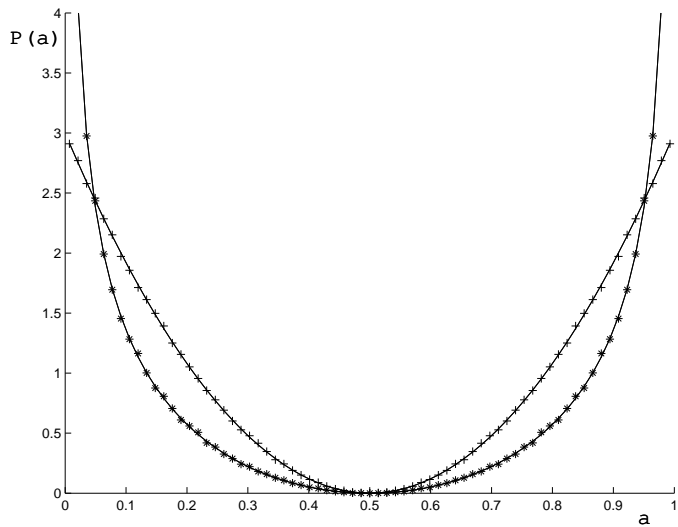}
\caption{Distribution  $P(a)$ of an eigenvalue $a$ of random
 matrices of size $N=2$ distributed according to the HS measure $(+)$
and the Bures measure $(*)$.
Solid lines refer to analytical predictions (\ref{dist2c}).
}
 \label{fig1}
\end{center}
 \end{figure}

\begin{figure} [htbp]
   \begin{center}
 \includegraphics[width=6.3cm,angle=0]{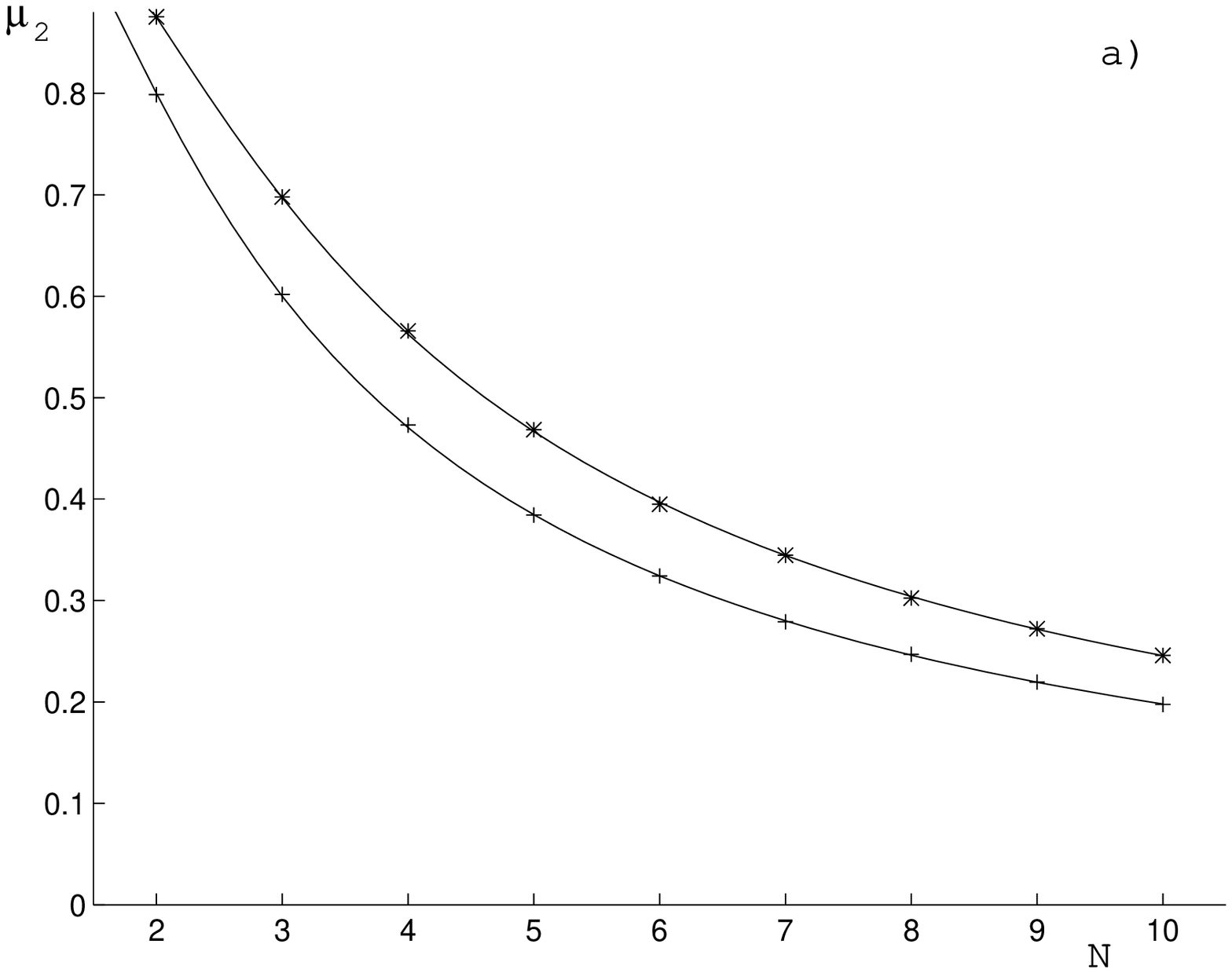}
 \includegraphics[width=6.3cm,angle=0]{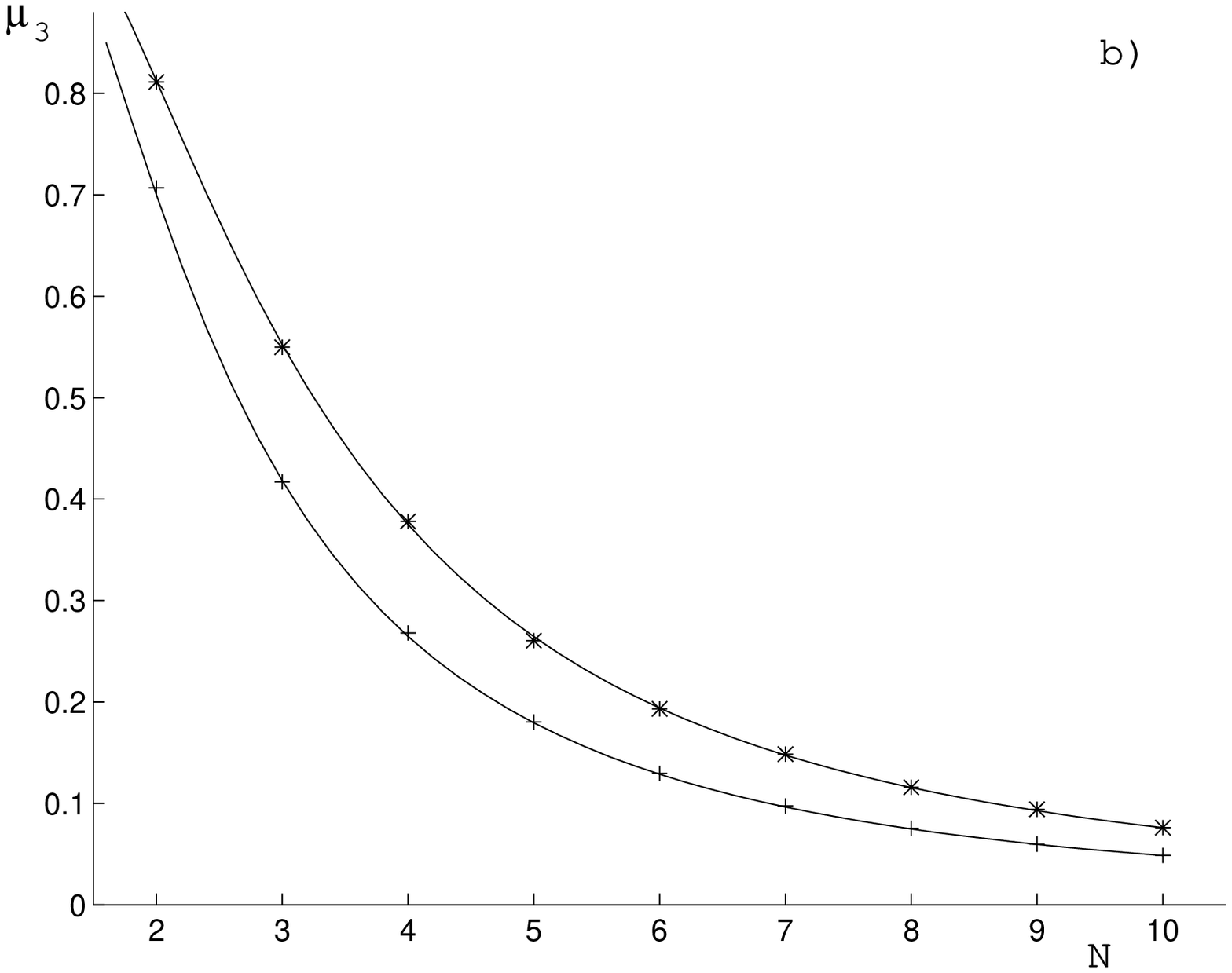}
\caption{Average traces  $\kappa_m=\langle {\rm Tr} \rho^m \rangle_B$
for a)  $m=2$ and b)  $m=3$ for an ensemble of Bures random density matrices
($\times$) and HS random states ($+$) of size  $N=2,\dots, 10$.
Solid lines represent interpolations of analytical results (\ref{momkbur}).
}
 \label{fig2}
\end{center}
 \end{figure}

To evaluate advantages of this algorithm in action
we have generated in this way several random Bures matrices
of different sizes.
In the one-qubit case, $N=2$
we  calculated the distribution of an eigenvalue
$a=\lambda_1=1-\lambda_2$
for the Hilbert--Schmidt and the Bures measures and compared in
Fig. \ref{fig1} our numerical data
with analytical results obtained in \cite{ZS01},
 \begin{equation}
\label{dist2c}
P_{\rm HS}(a)= 12(a-1/2)^2 , \quad \quad
P_B(a)= \frac{ 8(a-1/2)^2}{ \pi \sqrt{a(1-a)}},
\end{equation}
where $a\in [0,1]$.
For larger dimensions we computed the mean traces
$\kappa_m=\langle {\rm Tr} \rho^m \rangle_B$
averaged over the Bures measure (\ref{dmuM}).
Fig. \ref{fig2} shows the comparison of the numerical data
with analytical results following from \cite{Ha98,SZ04},
 \begin{equation}
\label{momkbur}
\langle {\rm Tr} \rho^2 \rangle_B =\frac{5N^2+1}{2N(N^2+2)},
\quad \quad
 \langle {\rm Tr} \rho^3 \rangle_B =
\frac{8N^2+7}{(N^2+2)(N^2+4)} .
\end{equation}
Observe that the average purity of the Bures states is higher
than the averages computed with respect to the HS measure.
This shows that the Bures measure is more concentrated
in the vicinity of the pure states than the flat measure.

Demonstrating practical usefulness of the algorithm
to generate random matrices according
to formula (\ref{burrand}) we may generalise it to
get a one-parameter family of
interpolating  ensembles of random matrices.
Taking any fixed parameter $x$ from the interval $[0,1]$
and setting $y=1-x$ we may construct a random density matrix
from a random Ginibre matrix $A$ and a random unitary matrix $U$,
\begin{equation}
\rho_{\rm x} \ = \  \frac{\ \ (y{\mathbbm 1}+xU)AA^{\dagger} (y{\mathbbm 1}+xU^{\dagger})}
{{\rm Tr}[(y{\mathbbm 1}+xU)AA^{\dagger} (y{\mathbbm 1}+xU^{\dagger})]} .
\label{burrandx}
\end{equation}
It is clear that for $x=0$ this expression
reduces to (\ref{hsrand}) and produces
a density matrix distributed according to the Hilbert-Schmidt measure,
while for $x=1/2$ one gets a Bures density matrix.
Since the Ginibre ensemble is invariant with respect to
unitary rotations, $A\to UAU^{\dagger}$,
increasing the value of $x$ above $1/2$
one interpolates back to the HS measure,
which is obtained again for $x=1$.
Note that the critical parameter $x_c$,
at which the transition between both ensembles
effectively takes place, is dimension dependent,
$x_c= x_c(N)$.

\section{Real Bures density matrices}\label{Sec4}

A similar construction can also be used to construct random real density matrices.
To generate these matrices according to induced measure \cite{ZS03}
\begin{equation}
\label{pmuhsr}
P_M^R(\rho)  \;  \propto \;
\Theta(\rho)\; \delta ( 1 - {\rm Tr} \rho)
 ({\rm det} \rho)^{(M-N-1)/2} \;
\end{equation}
one uses the same formula (\ref{hsrand}) with a random matrix $A$
of the real Ginibre ensemble. Note that in the case of real density matrices
the Hilbert--Schmidt measure is obtained for $M=N+1$,
since in this case the last factor is equal to unity.
For this end one needs to generate a
rectangular real Ginibre matrix $A$
of dimension $N \times (N+1)$.

To obtain real Bures matrices we begin with an analog of
eq.(\ref{pbur1}) in which $A$ is a real Ginibre matrix of size $N \times M$,
while $H$ stands for a real symmetric matrix of size $N$,
\begin{equation}
\label{pbure1}
P_B(\rho) \propto \int dA \int dH e^{-{\rm Tr}[AA^T+ H^2 AA^T]}
\ \delta \left(  \rho - \frac{AA^T}{{\rm Tr} AA^T}\right) .
\end{equation}
As in the complex case we introduce the $\delta$--function
by integrating over $s$
and rescale both matrices $A$ and $H$ to obtain expressions
\begin{eqnarray*}
\label{pbure3}
\fl P^R_B(\rho) & \propto \int_0^{\infty}
\frac{d s}{s} e^{-s} s^{MN/2}
 \int dH e^{- s {\rm Tr}H^2 \rho}
\int dA \delta ( \rho - {\rm Tr} AA^T)\;
\delta ( 1 - {\rm Tr} \rho) \nonumber \\\fl
        &  \propto \int_0^{\infty}
\frac{d s}{s} e^{-s} s^{MN/2-N(N+1)/4}
\prod_i  \frac{1}{\sqrt{\lambda_i}}
\prod_{i <j}   \frac{1}{\lambda_i+\lambda_j}
\Theta(\rho)  \delta ( 1 - {\rm Tr} \rho)
({\rm det} \rho)^{(M-N-1)/2} .
\end{eqnarray*}
This expression coincides with the real Bures measure for
$M=N+1$. In this case we perform now
 another rescaling, $A \to \frac{1}{\sqrt{{\mathbbm 1}+H^2}}A$.
and apply lemma 2 from \ref{app1}.
In this way we replace an integral over symmetric matrices $dH$
by an integral over the measure d$\mu_o$ on symmetric
unitary matrices, characteristic of circular orthogonal ensemble (COE).
The final expression
\begin{equation}
\label{pbure6}
P_B(\rho) \ \propto \ \int_{U(N)} d\mu_o(U)
\int dA e^{-{\rm Tr}AA^T}
\ \delta \left( \rho - \frac{   \frac{|{\mathbbm 1}+U|}{2}
AA^{\dagger} \frac{|{\mathbbm 1}+U^{\dagger}|}{2}}
{{\rm Tr}   \frac{|{\mathbbm 1}+U|}{2}
AA^{\dagger} \frac{|{\mathbbm 1}+U^{\dagger}|}{2}}
 \right) .
\end{equation}
allows us to write down the final expression
for a real random Bures matrix
\begin{equation}
\rho^R_{\rm B} \ = \  \frac{\ \ |{\mathbbm 1}+V|AA^T
|{\mathbbm 1}+V^{\dagger}|}
{{\rm Tr}|{\mathbbm 1}+V|AA^T |{\mathbbm 1}+V^{\dagger}|} \,
\label{burrerand}
\end{equation}
Here $|X|$ denotes $\sqrt{XX^{\dagger}}$, while
$A$ represents a real rectangular random
Ginibre matrix of dimension $N \times (N+1)$,
and $V$ is a unitary matrix from the ensemble
of symmetric unitary matrices (COE).
To generate such a symmetric matrix one may take any matrix $U$
distributed according to the Haar measure on $U(N)$
and set $V=UU^T$ \cite{Me91}.
Also in the real case one may design
a one parameter ensemble analogous to  (\ref{burrandx}),
which interpolates between
the HS and Bures measures.
\medskip

\begin{figure} [htbp]
   \begin{center}
 \includegraphics[width=9.0cm,angle=0]{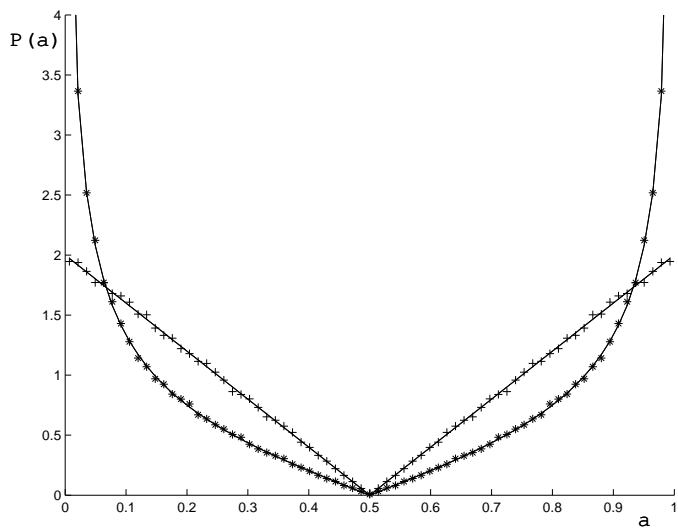}
\caption{As in Fig. \ref{fig1} for real random matrices.
Numerical data agree with analytical predictions (\ref{dist2r}).
}
 \label{fig3}
\end{center}
 \end{figure}

After generating numerically several real random Bures density matrices
we analysed their spectra.
In Fig. \ref{fig3} we compare
the distribution of an eigenvalue $P(a)$
of a real one-qubit random  state
for the HS and Bures measures with the
corresponding analytical results \cite{SZ03,ZS03},
\begin{equation}
\label{dist2r}
P^R_{\rm HS}(a)= 4|a-1/2| , \quad \quad
P^R_B(a)= \frac{  |a-1/2|}{\sqrt{a (1-a)}} .
\end{equation}

 \section{Moment Generating function}\label{Sec5}
We are going to analyse the
moments of a homogeneous function $F_q(\lambda)$
of the eigenvalues $\lambda_i$ of degree $q$
for random matrices distributed with respect to the induced measure (\ref{dmuM})
and  the Bures measure (\ref{dmuB}).
It is convenient to consider the corresponding Laguerre ensembles:
\begin{equation}
\label{dmuML} d\mu_M^L(\lambda_1,...\lambda_N) \propto  \exp({-\sum_i\lambda_i})
\prod_{i<j}^{1...N} (\lambda_i -\lambda_j )^2 \prod_i \Theta(\lambda_i) \lambda_i^{M-N} {\rm d}\lambda_i
\end{equation}
and
\begin{equation}
\label{dmuBL} d\mu_B^L(\lambda_1,...\lambda_N) \propto \exp(-\sum_i\lambda_i  )
 \prod_{i<j}^{1...N} {(\lambda_i -\lambda_j )^2 \over \lambda_i +\lambda_j}\prod_i \Theta(\lambda_i)
 \lambda_i^{-1/2} {\rm d}\lambda_i\ .
\end{equation}
The reason is that the moments and the averages are closely related:
 \begin{equation}\label{FqM}
       \langle F_q(\lambda)\rangle_M =\int {\rm d} \mu_M F_q(\lambda)={\Gamma(MN)\over \Gamma(MN +q)}
\langle F_q(\lambda)\rangle_M^L
\end{equation}
and likewise
 \begin{equation}\label{FqB}
       \langle F_q(\lambda)\rangle_B =\int {\rm d} \mu_B F_q(\lambda)={\Gamma(N^2/2)\over \Gamma(N^2/2 +q)}
\langle F_q(\lambda)\rangle_B^L
\end{equation}
where $\Gamma(x)$ is Euler's Gamma function, which can simply be proven by going to spherical coordinates.
Thus we have the relations for the moments of purity
 \begin{equation}\label{PrM}
       \langle P^r\rangle_M ={\Gamma(MN)\over \Gamma(MN +2r)}\langle P^r\rangle_M^L
\end{equation}
and
\begin{equation}\label{PrB}
       \langle P^r\rangle_B ={\Gamma(N^2/2)\over \Gamma(N^2/2 +2r)}\langle P^r\rangle_B^L\ .
\end{equation}
In the same way we may introduce the matrix Laguerre ensembles ${\rm d}\mu_M^L(\rho)$ and
 ${\rm d}\mu_B^L(\rho)$. Then we consider the matrix Laplace transforms of these ensembles
\begin{equation}\label{KrM}
        \int \exp(-\Tr K\rho) {\rm d}\mu_M^L(\rho)= \prod_{i=1}^N {1\over (1+K_i)^M}
\end{equation}
and
\begin{equation}\label{KrB}
        \int \exp(-\Tr K\rho) {\rm d}\mu_B^L(\rho)= \prod_{ij}^{1...N} {2\over \sqrt{1+K_i} +\sqrt{1+K_j}}
\end{equation}
which have been calculated elsewhere~\cite{SZ04}, $K$ is Hermitian $K\ge 0$ with eigenvalues $K_i$.
From these we can derive the generating functions for the moments of purity
\begin{equation}\label{ZML}
        Z_M^L(x)=\int  {\rm e}^{-x\Tr \rho^2}{\rm d}\mu_M^L(\rho)={\rm e}^{-x\Tr(\delta/\delta\bm K )^2}
 \prod_{i=1}^N {1\over (1+K_i)^M} \biggr \vert_{\bm K=\bm 0}\end{equation}
and similarly
\begin{equation}\label{ZBL}
        Z_B^L(x)= {\rm e}^{-x\Tr(\delta/\delta\bm K )^2} \prod_{ij}^{1...N} {2\over \sqrt{1+K_i} +\sqrt{1+K_j}}
 \biggr \vert_{\bm K=\bm 0} \ .
\end{equation}

Applying the matrix differential operator $\Tr(\delta/\delta K )^2$ on some
invariant function it can be
 expressed in eigenvalues $K_i$ using the Vandermonde determinant $ \Delta(\bm K)=\prod_{i<j} (K_i-K_j)$
\begin{equation}\label{TddK}
      \Tr(\delta/\delta\bm K )^2=\Delta(\bm K)^{-1}\sum_i\left ({\partial\over \partial K_i}\right )^2\Delta(\bm K)\
\end{equation}
It is easily seen that this operator is Hermitian.
Calculation of all the derivatives, which are needed, is not so simple.
Instead we make a Hubbard--Stratonovich transformation of the exponential operator acting on some invariant function $F(K)$ \ of degree $q$,
\begin{eqnarray}  \nonumber
   \fl   {\rm e}^{-x\Tr(\delta/\delta K )^2} F(K)= \int {\rm D}Y
\exp(-\Tr Y^2)\ \exp(2i \sqrt{x} \Tr Y {\delta\over \delta\bm K })
F(\bm K) \biggr|_{\bm K=\bm 0} \\
    \ \ \ \ \ = \int {\rm D}Y \exp(-\Tr Y^2) F(2i \sqrt{x} Y )\ . \label{eTF}
\end{eqnarray}
Here $Y$ is a Hermitian matrix. Thus we have reduced this expression to an average over the Gaussian unitary ensemble (GUE).
We choose the normalisation condition $\int {\rm D}Y \exp(-\Tr Y^2)=1$.
For the generating functions we obtain
\begin{equation}\label{ZML1}
        Z_M^L(x)= \int {\rm D}Y \exp(-\Tr Y^2)  \prod_{j=1}^N {1\over (1+2i\sqrt{x} Y_j)^M}  \end{equation}
and similarly
\begin{equation}\label{ZBL1}
        Z_B^L(x)=  \int {\rm D}Y \exp(-\Tr Y^2) \prod_{jk}^{1...N} {2\over \sqrt{1+2i\sqrt{x} Y_j} +\sqrt{1+2i\sqrt{x} Y_k}}\ .
\end{equation}
Thus $Z_M^L(x)$ is related to some negative moment of the characteristic polynomial in GUE, while $Z_B^L(x)$ is
something more complicated - nevertheless also written as some GUE average.
The above expressions may be considered as a starting point
for calculation of the moments of the distribution of purity.
Since such results were already obtained by Giraud
for random matrices distributed with respect to HS measure \cite{Gi07},
we will not discuss this case any further, but we shall rather concentrate
on the more complicated case of random states distributed with
respect to the Bures measure.

Let us write $Z_B^L(x)$
as an integral over eigenvalues $Y_i$ of $Y$ (with the constant $C_N=\frac{2^{N(N-1)/2}}{\pi^{N/2}\prod_{j=1}^{N+1}\Gamma(j)}$ normalising the Gaussian measure)
\begin{eqnarray}  \nonumber
   \fl Z_B^L(x)=  C_N \int\prod_i{\rm d}Y_i{\rm e}^{-\sum_iY_i^2} \prod_{i<j}^{1...N} (Y_i-Y_j)^2 \prod_{jk}^{1...N}
{2\over \sqrt{1+2i\sqrt{x} Y_j} +\sqrt{1+2i\sqrt{x} Y_k}}\\\fl  ={C\over(i\sqrt{x})^{N(N-1)}}
\int\prod_i{{\rm d}Y_i\over \sqrt{1+2i\sqrt{x}Y_i}}{\rm e}^{-\sum_iY_i^2}
 \prod_{i<j}(( \sqrt{1+2i\sqrt{x}Y_i}-\sqrt{1+2i\sqrt{x}Y_j}))^2 \ .
\label{ZBL2}
\end{eqnarray}
\begin{figure}[htbp]
   \begin{center}
			\psset{unit=.5mm,linecolor=black,linewidth=1}
			\begin{pspicture}(290,100)
				\psline[linewidth=1.]{->}(30,10)(30,90)
				\psline[linewidth=1.]{->}(10,50)(110,50)
				\psline[linewidth=0.4,linestyle=dashed,dash=6. 4.](10,40)(110,90)
				\psline[linewidth=0.4,linestyle=dashed,dash=6. 4.](10,60)(110,10)
				\psbezier(110,88)(33,50)(33,50)(110,12)
				\pscircle[fillstyle=solid,linewidth=0.5](49,17){7}
				\put(45,15){\normalfont\normalsize $\bm {z_i}$}
				\pscircle[fillstyle=solid,fillcolor=black,linewidth=0.3](52.5,50){1.5}
				\put(47,51){\normalfont\footnotesize $\bm 1$}
				\put(90,70){\normalfont\normalsize $\cal{C}$}
				\psarc[linewidth=0.4]{-}(30,50){13}{155}{180}
				\psarc[linewidth=0.4]{->}(30,50){13}{120}{155}
				\put(16,63){\normalfont\footnotesize $\frac{\pi}{4}$}
				\psarc[linewidth=0.4]{-}(30,50){17}{180}{205}
				\psarc[linewidth=0.4]{<-}(30,50){17}{205}{230}
				\put(13,30){\normalfont\footnotesize $\frac{\pi}{4}$}
				\psline[linewidth=1.]{->}(200,10)(200,90)
				\psline[linewidth=1.]{->}(160,50)(280,50)
				\psline[linewidth=0.4,linestyle=dashed,dash=6. 4.](160,20)(280,80)
				\psline[linewidth=0.4,linestyle=dashed,dash=6. 4.](160,80)(280,20)
				\psbezier(160,81)(220,53)(220,53)(280,81)
				\psframe[fillstyle=solid,linewidth=0.5,framearc=0.5](177,10)(245,24)
				\put(180,15){\normalfont\normalsize $\bm {z_i\to 1+i\sqrt{x} z_i}$}
				\pscircle[fillstyle=solid,fillcolor=black,linewidth=0.3](220,50){1.2}
				\put(217.5,51.5){\normalfont\footnotesize $\bm 1$}
				\put(260,75){\normalfont\normalsize $\cal{C'}$}
			\end{pspicture}

\caption{Contour of integration, $\cal{C}$, in eq.(\ref{ZBL3}) and its transformation into the contour $\cal{C'}$, right hand side picture, according to the rule $z_i\to1+i\sqrt{x} z_i$.}
 \label{fig4}
\end{center}
 \end{figure}
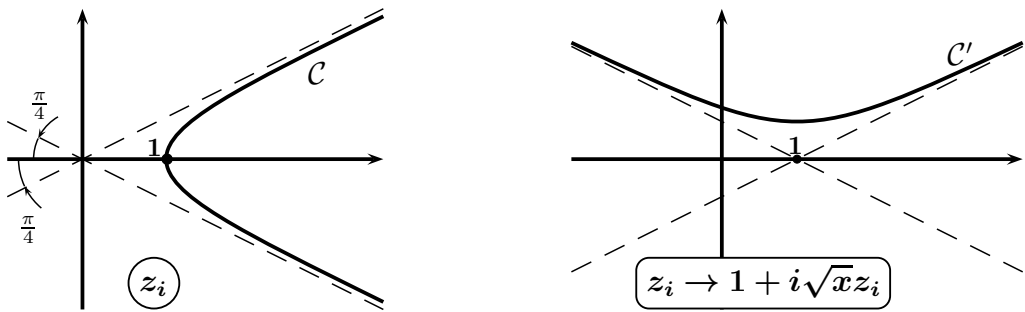Here the $Y_i$ integrations run from $-\infty$ to $+\infty$. Introducing the complex integration variables
 $ z_j= \sqrt{1+2i\sqrt{x}Y_j}$ with $z_j {\rm d}z_j = i \sqrt{x} {\rm d}Y_j$ we obtain
\begin{equation}\label{ZBL3}
   Z_B^L(x)  ={C_N\over(i\sqrt{x})^{N^2}}\int_{\cal C ^N}\prod_i{\rm d}z_i\exp\left( {1\over 4x}(z_i^2-1)^2\right)\cdot \Delta_N^2(\bm z)\ .
         \end{equation}
where all the $z_i$ run over a contour $\cal C$ starting from ${\rm e}^{-i\pi/4}\cdot \infty$ to ${\rm e}^{+i\pi/4}\cdot \infty$ passing through the
saddle $z_i=1$ (see fig.\ref{fig4}). We want to expand $Z_B^L(x)$ in powers of $-x$ to obtain the averaged
moments $\langle P^r\rangle_B^L/r!$. This means that we have to do saddle-point integration for $1/x \to \infty$.
The relevant saddle point is $z_i=1$. Now we want to expand around the saddle point and make the
transformation $z_i \to 1+i\sqrt{x} z_i$:
\begin{equation}\label{ZBL4}
   Z_B^L(x)  ={C_N  }\int_{\cal {C'} ^N}\prod_i{\rm d}z_i \exp(-z_i^2-i\sqrt{x} z_i^3+x z_i^4/4) \cdot \Delta_N^2(\bm z)\ .
         \end{equation}
The new contour  $\cal C'$ (see fig.\ref{fig4}) is such that the integral converges.

Now we may expand in powers of $\sqrt{x}$, which turns out to become a power series in $x$. In each term we may deform the
contour back to the real axis and thus obtain, at least for the asymptotic expansion for $x\to 0$, the same expansion as for
\begin{equation}\label{ZBL5}
   Z_B^L(-x)  ={C_N  }\int_{\cal {R} ^N}\prod_i{\rm d}z_i\exp(-z_i^2+\sqrt{x} z_i^3-x z_i^4/4) \cdot \Delta_N^2(\bm z)\ .
\end{equation}
with $x>0$. Thus
\begin{equation}\label{Z_def}
 Z_B^L(-x)=\sum_{r=0}^{\infty} x^r\langle P^r\rangle_B^L .
\end{equation}
The next section is devoted to the derivation of analytic expressions for moments $\langle P^r\rangle_B^L$.

\section{Derivation of moments of purity for Bures measure}\label{Sec6}
\label{part6}
Coefficients of the Taylor expansion of $Z_B^L(-x)$ in the vicinity of $x=0$ are nothing but, up to a constant, moments of traces $\tr \bm  z^3$, $\tr \bm  z^4$ and their powers averaged with respect to the probability measure corresponding to GUE. Therefore, our main interest here is in calculation of the quantities that below are referred to as $T_{k,m}$,
\begin{equation}\label{moments_def}
 T_{k,m}=\left\langle\big(\tr\bm  z^4\big)^k\big(\tr\bm  z^3\big)^{2m}\right\rangle_{GUE_{N\times N}}\;,\quad T_{0,0}\equiv 1.
\end{equation}
The connection between $\langle P^r\rangle_B^L$ and $T_{k,m}$'s is given by the formula
\begin{equation}\label{moments_rel}
\langle P^r\rangle_B^L=\sum_{m=0}^r\frac{(-1)^{r-m}2^{2(m-r)}r!}{(r-m)!(2m)!}T_{r-m\;m}.
\end{equation}
See explanations under the formula~(\ref{J_expansion})

Calculation of $T_{k,m}$ for general $k$ and $m$ is a rather nontrivial problem. To do this we are going to derive a system of recurrency relations that will allow us, in principle, to obtain a closed form of $T_{k,m}$ for all particular values of $k$ and $m$. In the basis of derivation of such a recurrency lays a so-called ``deform-and-study'' approach, a string theory technique of revealing hidden symmetries. For the first time, this technique as a closed calculation method to the problems of Random Matrix Theory appeared in the work~\cite{AvM} by Adler and van Moerbeke where it was utilised to study gap-formation-probability integrals over various Unitary Ensembles. Later, this approach was modified to calculate distribution properties (such as cumulants) in other random matrix models~\cite{OK}.

The celebrated result of the theory of integrable system states that internal symmetry of matrix integrals of $\beta=2$ Dyson's class (encoded in squared Vandermonde determinant) leads to highly non-trivial nonlinear relations between combinations of averaged traces. One of them which is of primary importance for the approach considered is the Kadomtsev-Petviashvili (KP) equation (see~\cite{AvM}).

The most economic way to work with these relations is to introduce a $\bm t$-deformation into the integration measure. Instead of an original matrix integral
\begin{equation}\label{I_original}
\mathcal{I}_N(x)=\int_{\cal{R}^N}\Delta_N^2(\bm z)\prod_i^{1\dots N} \dm{ z_i;x}
\end{equation}
with some given measure $\dm{ z;x}$ depending on a ``physical'' parameter $x$, one considers the integral depending on an infinite set of auxiliary parameters $t_k$, $k=0,1,\dots$
\begin{equation}\label{tau_general}
\tau_N\set{\bm t}\equiv\frac{1}{N!}\int_{\cal{R}^N}\Delta_N^2(\bm z)\exp\left\lbrace \sum_{k=0}^\infty t_k\tr\bm z^k\right\rbrace \prod_{i} \dm{ z_i;x}\;.
\end{equation}
The KP-equation being written down in the variables $\bm t$ has the following form
\begin{equation}\label{KP}
\left(\frac{\partial^4}{\partial t_1^4}+3\frac{\partial^2}{\partial t_2^2}-4\frac{\partial^2}{\partial t_1\partial t_3} \right) \log\tau_N+6\left( \frac{\partial^2}{\partial t_1^2}\log\tau_N\right)^2 =0.
\end{equation}

In the theory of integrable systems such objects as defined in~(\ref{tau_general}) are referred to as $\tau$-functions. Note that $\tau_N$ depends on an infinite set of parameters and satisfies an infinite set of relations, one of those is given by~(\ref{KP}). Strictly speaking, one has to restrict the number of parameters $t_k$ and then attach correct signs to each of them in order to attain convergence of the integral. Since eventually all of them are set to zero, we do not need to pay them any special attention.

The differential equation for the function $\mathcal{I}_N(x)$ is obtained by projecting the KP-equation onto the hyperplane $\bm t=\bm 0$. To perform this projection Adler, Shiota and van Moerbeke~\cite{AvM} suggested to use Virasoro constraints (VC) as an additional block giving a link between the $t_k$-derivatives in~(\ref{KP}) and the derivatives over $x$ that supplemented the deform-and-study approach to a complete tool applicable to calculation of random matrix integrals. On the contrary to the KP-equation, the particular form of the VC as well as the way of their derivation is essentially influenced by the choice of the measure $\dm{ z;x}$. The basic idea of derivation can be expressed in the following simple form
$$
\frac{\delta}{\delta\eps}
	\left(
		\tau_N\set{\bm t}
			\Big|_{ z_j\mapsto z_j+\delta\eps\; z_j^{q+1}\mathrm{Poly}( z_j)}
	\right)\equiv0\;,\quad q=-1,0,1,\dots
$$
while the details may vary from one model to the other, and they can be demonstrated much simpler by studying of our particular examples.

The final step of the approach is to resolve the obtained VC and KP-equation jointly on the hyperplane $\bm t=\bm 0$ to bring a closed equation for $\mathcal{I}_N(x)$. Substitution of $\mathcal{I}_N(x)$ in the form of a Taylor series into the obtained equation gives rise to a recurrence relation for the coefficients of the series and subsequently for the sought moments $T_{k,m}$, eq.~(\ref{moments_def}).

Below we show how this approach can be applied to calculate the moments $T_{k,m}$ defined by~(\ref{moments_def}). All details of calculations are given in two appendices (\ref{app2} and~\ref{app3}), below we give only a plan of the calculation program.

We start with the derivation of the recurrence relation for $T_{k,0}$. This choice is dictated by two reasons, first, by the relative simplicity of this case and, second, because the moments $T_{k,0}$ serve as an initial condition for calculation of the higher order moments.

To derive expressions for the moments $T_{k,0}=\left\langle\big(\tr\bm  z^4\big)^k\right\rangle_{GUE_{N\times N}}$ we consider an auxiliary integral, $\mathcal{J}_N(x)$,
\begin{equation}\label{J_aux}
 \mathcal{J}_N(x)=\frac{2^{N(N-1)/2}}{\pi^{N/2}\prod_{j=1}^{N+1}\Gamma(j)}\int_{\cal{R}^N}\Delta_N^2(\bm z)
	\prod_{j}^{1\dots N} e^{- z_j^2-x z_j^4}d z_j.
\end{equation}
Obviously, the sought moments can be obtained as coefficients in the expansion of this integral into the Taylor series in the vicinity of $x=0$. Thus, deriving a differential equation on $\mathcal{J}_N(x)$ by using the method discussed in the main section enables us to link the moments by a recurrence relation.

Direct application of the ``deform-and-study'' approach to the integral~(\ref{J_aux}) does not help in the derivation of a differential equation for the function $\mathcal{J}_N(x)$ (this is discussed in details in~\ref{app2}).  However, due to its symmetry the integral $\mathcal{J}_N(x)$ can be represented as a product of two simpler integrals:
\begin{eqnarray}\label{G_apear}
\mathcal{J}_N(x)=\frac{2^{N(N-1)/2}}{\pi^{N/2}\prod_{j=1}^{N}\Gamma(j)}
	\cases{
			\Xi_k^+(x)\Xi_k^-(x)\;,& $N=2k$;\\
			\Xi_k^+(x)\Xi_{k+1}^-(x)\;,& $N=2k+1$;
                }
\end{eqnarray}
where
\begin{equation}\label{G_def}
\Xi_k^\nu(x)=\frac{1}{k!}\int_{{\cal R}_+^k}\Delta_k^2(\bm z)
	\prod_{j}^{1\dots k} z_j^{\nu/2}e^{- z_j-x z_j^2}d z_j,\qquad \nu=\pm1.
\end{equation}
Integral $\Xi_k^\nu(x)$ can be investigated with the help of the announced approach. Details of calculations and results are given in~\ref{app2}. Here, however, we make an important remark. There is another way to derive equation~(\ref{ApD}) from the appendix. It is a consequence of the known result by P.J. Forrester and N.S. Witte~\cite{FW}. They showed by using other methods that the matrix integral (here we use their original notation for integral)
$$
\tilde{E}_n(s;\nu)=\frac{1}{n!}\int_{(-\infty,s]^n}\Delta_n^2(\bm z)\prod_{j=1}^n(s- z_j)^\nu e^{- z_j^2}d z_j
$$
is expressed in terms of a solution of the Painlev\'{e}~IV equation. Namely,
\begin{equation}\label{PIV1}
\frac{d}{ds}\log\tilde{E}_n(s;\nu)=\phi(s)\;,
\end{equation}
where $\phi(s)$ satisfies the differential equation, which is the Painlev\'{e}~IV equation written down in the Chazy form
\begin{equation}\label{PIV}
\phi'''+6(\phi')^2+4[2(n-\nu)-s^2]\phi'+4s\phi-8n\nu=0.
\end{equation}
Then, due to the relation
$$
\tilde{E}_k\left( -\frac{1}{2\sqrt{x}}\;;\frac{\nu}{2}\right) =x^{\frac{k(2 k+\nu)}{4}}e^{-\frac{k}{4x}}\Xi_k^\nu(x)
$$
one can restore~(\ref{ApD}) from~(\ref{PIV}) and~(\ref{PIV1}) by appropriate change of variables. In spite of the fundamental character of the obtained Painlev\'{e}, the equation~(\ref{ApD}) derived by our regular method is more convenient for the analysis that is done in~\ref{app2}.

To derive the recurrence relation for the $T_{k,m}$ with the help of ``deform-and-study'' approach as well as in the previous case we define an auxiliary integral
\begin{equation}\label{J2_aux}
 \mathcal{J}_N(x,y)=\frac{2^{N(N-1)/2}}{\pi^{N/2}\prod_{j=1}^{N+1}\Gamma(j)}\int_{{\cal R}^N}\Delta_N^2(\bm z)
	\prod_{j=1}^N e^{- z_j^2+y z_j^3-x z_j^4}d z_j.
\end{equation}
To make the further procedure of derivation successful we have introduced the extra parameter~$y$. Appearance of more then one variable apparently lead to a differential equation in partial derivatives in both variables.

The form of integral~(\ref{J2_aux}) implies that one can seek the solution of the obtained equation in the form of a series in~$x$ and~$y$:
\begin{equation}\label{J_expansion}
 \mathcal{J}_N(x,y)=\sum_{k,m=0}^\infty(-1)^kT_{k,m}\frac{y^m x^k}{m!k!}.
\end{equation}
Expansion~(\ref{J_expansion}) is used to derive relation~(\ref{moments_rel}) between averaged moments $\langle P^r\rangle_B^L$ and $T_{k\;m}$. Indeed, it is enough to compare coefficients standing at equal powers of $x$ in the Taylor expansion of the both sides of the identity $\mathcal{J}_N(x/4,\sqrt{x})= Z_B^L(-x)$. Expansion in the left hand side can be obtained from eq.~(\ref{J_expansion}) after the change $x\to x/4$; $y\to \sqrt{x}$, while in the right hand side one uses the Taylor expansion~(\ref{Z_def}).

Explicit expressions for the first several moments $T_{k,m}$ are given in~\ref{app3}. The higher moments of purity follow from the general expression~(\ref{PrB}) which being combined with the relation~(\ref{moments_rel}) gives
\begin{equation}
\label{mukk}
\mu_r= \langle ({\rm Tr} \rho^2)^r\rangle_B = \frac{\Gamma(N^2/2)}{\Gamma(N^2/2+2r)}\sum_{m=0}^r\frac{(-1)^{r-m}2^{2(m-r)}r!}{(r-m)!(2m)!}T_{r-m\;m}.
\end{equation}

\begin{figure} [htbp]
   \begin{center}
 \includegraphics[width=9.0cm,angle=0]{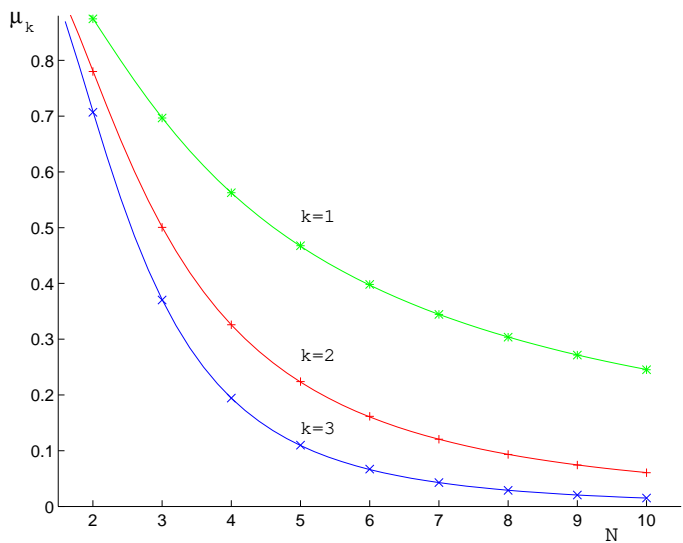}
\caption{Average moments of purity  $\mu_k=\langle ({\rm Tr} \rho^2)^k \rangle_B$
with $k=1,2,3$  for an ensemble of Bures random density matrices
of size  $N=2,\dots, 10$.
Solid lines represent interpolations of analytical results (\ref{purbur}).
}
 \label{fig5}
\end{center}
 \end{figure}

To demonstrate the ability of our approach we calculated the first three moments $\mu_r$ explicitly. The first moment $\mu_1$ coincides with the mean trace $\langle {\rm Tr} \rho^2 \rangle_B$ given by eq.~(\ref{momkbur}). Expressions for other two moments are reproduced below
 \begin{eqnarray}\label{purbur}
\mu_2 & = \langle ({\rm Tr} \rho^2)^2 \rangle_B =
\frac{5(5N^4+47N^2+32)} {2(N^2+2)(N^2+4)(N^2+6)}
\nonumber \\
\mu_3 & = \langle ({\rm Tr} \rho^2)^3 \rangle_B =
\frac{5(25N^8+690N^6+6015N^4+8750N^2+1152)}
{8N(N^2+2)(N^2+4)\cdots (N^2+10)} .\label{purbur1}
\end{eqnarray}
As shown in Fig. \ref{fig5} our numerical data obtained by generating
random Bures states according to the method presented in Sec.~\ref{Sec3}
coincide with the above analytical results.

\section{Concluding remarks}

In this work we proposed an explicit construction to
generate random density matrices according to the Bures measure.
A single complex random Bures density matrix
of size $N$ is obtained directly
out of one  complex random square matrix $A$ from the Ginibre ensemble and
one random Haar unitary matrix $U$ of size $N$.
Similarly, to generate a real Bures state of size $N$,
it is sufficient to have an rectangular, $N \times (N+1)$ real
Ginibre matrix and a random symmetric unitary matrix $V$.
These practical recipes are not only simple but also economic
and allow one to form the Bures prior, useful as an initial step
to apply the quantum Bayesian approach.

Studying the distribution of random Bures states
we have analytically determined the moments of their purity.
These results, derived by means of the theory of
integrable systems, reveal the power and usefulness
of this analytic technique. Note, that the Painlev\'e IV transcendent appears in a natural way as an intermediate step of our calculation.

\section*{Acknowledgements}
It is a pleasure to thank R. Blume-Kohout
 for several discussions on possible applications
of random Bures states.
Financial support by the Transregio-12 project
der Deutschen Forschungsgemeinschaft
the special grant number DFG-SFB/38/2007 of
Polish Ministry of Science and Higher Education
 and the European Research Project  COCOS is gratefully acknowledged.

\appendix
\section{Measures on Hermitian and unitary matrices}
\label{app1}

In this section we prove two lemmas which allow us to
replace integration over the set of Hermitian matrices
by integration over unitary matrices
with respect to the circular unitary ensemble (CUE)
and circular orthogonal ensemble (COE), respectively.

\medskip

{\bf Lemma 1.} {\sl The measure
$d H / [{\rm det}({\mathbbm 1}+H^2)]^N$
on the set of Hermitian matrices $H=H^{\dagger}$ of order $N$,
is equivalent to the Haar measure on $U(N)$,
which corresponds to CUE.}
\medskip

{\bf Proof.} We start with a substitution,
$U=({\mathbbm 1} - iH)/ ({\mathbbm 1} + iH)= 2/({\mathbbm 1} + iH)-2$.
Thus the measure $dU=-\frac{2}{{\mathbbm 1} + iH} i dH \frac{2}{{\mathbbm 1} + iH}$
implies that
$U^{-1}dU=-\frac{2}{{\mathbbm 1} - iH} i dH \frac{2}{{\mathbbm 1} + iH}$.
This allows us to write down an explicit expression for the metric
\begin{equation}
\label{met1}
(ds)^2 = {\rm Tr} (U^{-1}dU) (U^{-1}dU)^{\dagger}=
 4 {\rm Tr}
\frac{1}{{\mathbbm 1} + H^2} dH \frac{1}{{\mathbbm 1} + H^2}dH ,
\end{equation}
which implies the measure
\begin{equation}
\label{met2}
d \mu (U) \; = \;  \frac{d H} {
[{\rm det}( \frac{{\mathbbm 1} + H^2}{2}) ]^N}
\end{equation}
and completes the proof.

\medskip

A similar lemma can be formulated for the
measure on $U(N)$ related to COE.
\medskip

{\bf Lemma 2.} {\sl The measure
$d H / [{\rm det}({\mathbbm 1}+H^2)]^{(N+1)/2}$
on the set of symmetric matrices $H=H^T=H^{\dagger}$ of order $N$,
is equivalent to the COE measure $\mu_o$ on $U(N)$.}

\medskip

Its proof is analogous to the previous one
and the exponent $(N+1)/2$
is related to the number of $N(N+1)/2$
of the independent variables of a random symmetric matrix.

\section{Recurrence relations for $T_{k,0}$}
\label{app2}

{\bf Proof of~(\ref{G_apear}).}
To prove~(\ref{G_apear}) we notice that the matrix integral $\mathcal{J}_N(x)$ has a determinantal representation:
$$
\mathcal{J}_N(x)\sim\det_{0\le i,j\le n-1}\left[\int_{\cal R} z^{i+j}e^{- z^2-x z^4}d z \right]
$$
Due to the symmetry of the function under integral the matrix of moments in the above determinant has a chessboard structure with zeros on all ``white squares'', i.e. for the elements with $i+j=2k+1$ ($k=0,1,\dots$). This type of determinants can be reduced by permutation of lines and rows to a determinant of block-diagonal matrix and as a result to a product of two determinants. In our particular case it gives
$$
\det_{0\le i,j\le N-1}[\mu_{i+j}]=\det_{0\le i,j\le \left\lfloor \frac{N}{2}\right\rfloor}[\mu_{2i+2j+2}]\det_{0\le i,j\le \left\lceil \frac{N}{2}\right\rceil}[\mu_{2i+2j}]=\det_{0\le i,j\le \left\lfloor \frac{N}{2}\right\rfloor}[\mu^+_{i+j}]\det_{0\le i,j\le \left\lceil \frac{N}{2}\right\rceil}[\mu^-_{i+j}]
$$
where $\left\lceil \bullet\right\rceil$ and $\left\lfloor \bullet\right\rfloor$ denote the integer part of a real number, so that
$
\left\lfloor x\right\rfloor\le x\le\left\lceil x\right\rceil
$ and the momentum matrix $\mu^\pm_{i+j}$ is obtained from $\mu_{2i+2j}$ and $\mu_{2i+2j+2}$ by the change of variables $ z^2\to z$  in corresponding integrals,
$$
\mu^\pm_{i+j}(x)=\int_{{\cal R}_+} z^{i+j\pm1/2}e^{- z-x z^2}d z
$$
Now returning back from the determinants to the matrix integral representation we obtain~(\ref{G_apear}).
\medskip

{\bf $\tau$-function and Virasoro constraints.} The integral $\Xi_k^\pm(x)$ corresponds to the $\tau$-function of the form
\begin{equation}\label{tau1}
\tilde{\tau}_k\set{\bm t}=\frac{1}{k!}\int_{{\cal R}_+^k}\Delta_k(\bm z)
	\prod_{j=1}^k z_j^{\nu/2} \exp\left[- z_j-x z_j^2+\sum_{\ell=1}^\infty t_\ell z_j^\ell\right] d z_j\;,
\end{equation}
here, the parameter $\nu$ stands for $\pm1$.

To derive the Virasoro constraints (VC) first one has to choose an appropriate change of variables. The general recipe says that this transformation must be chosen as
$$
 z_j\mapsto z_j+\delta\eps z_j^{q+1} f( z_j)\prod_{k}( z_j-a_k)\;,\quad q=-1,0,1\dots
$$
where $a_k$ are the boundaries of integration domain without both infinities (if they are presented) and with excluded zeros of the polynomial function $f( z)$. The function $f( z)$ is, in turn, related to the original integration measure through the parametrisation
$$
\frac{dV( z)}{d z}=\frac{h( z)}{f( z)}\;,\quad\mbox{with}\quad h( z)=\sum_{k=0}^\infty g_k z^k\;,\quad f( z)=\sum_{k=0}^\infty f_k z^k
$$
where $V( z)$ is a confinement potential, in our case
$$
V( z)= z+x z^2-\frac{\nu}{2}\log z\;,\quad\frac{dV( z)}{d z}=\frac{2 z+4x z^2-\nu}{2 z}\;,
$$
and correspondingly $f( z)=2 z$. Since the only zero of $f( z)$ coincides with the only finite boundary point of the integration domain, $ z=0$, we should use the shift of the form
\begin{equation}\label{shift}
 z_j\to  z_j+2\delta\eps z_j^{q+2}\qquad q=-1,0,1\dots
\end{equation}
Substitution of this change of variables into the integral~(\ref{tau1}) and variation over $\delta\eps$ give rise to an infinite number of VC
$$
2\sum_{m=0}^{q+1}\frac{\partial\tilde{\tau}_k}{\partial t_m\partial t_{m+q+1}}+2\sum_{m=1}^\infty mt_m\frac{\partial\tilde{\tau}_k}{\partial t_{q+m}}+\nu\frac{\partial\tilde{\tau}_k}{\partial t_{q+1}}-2\frac{\partial\tilde{\tau}_k}{\partial t_{q+2}}-4x\frac{\partial\tilde{\tau}_k}{\partial t_{q+3}}=0
$$
where the first term originated from the squared Vandermonde determinant and the volume element $\prod_{j=1}^nd z_j$, the second term corresponds to $\bm t$-deformation, and the other three are the contributions from the measure. Note that the operation of differentiation over $t_0$ reduces to multiplication by the integral dimension, $\frac{\partial\tilde{\tau}_k}{\partial t_0}\equiv k$.
\medskip

{\bf Projection of KP onto the hyperplane $\bm t=\bm 0$.} To perform the projection of the KP-equation~(\ref{KP}) onto the hyperplane $\bm t=0$ one needs to know the following derivatives
$$
 \left.\frac{\partial^4 \log\tilde{\tau}_k}{\partial t_1^4}\right|_{\bm t=0}\;,\quad\left.\frac{\partial^2 \log\tilde{\tau}_k}{\partial t_1^2}\right|_{\bm t=0}\;,\quad\left.\frac{\partial^2\log\tilde{\tau}_k}{\partial t_1\partial t_3}\right|_{\bm t=0},
$$
the expression for the second derivative over $t_2$ immediately follows from the observation
$$
 \frac{\partial\tilde{\tau}_k}{\partial x}=-\frac{\partial\tilde{\tau}_k}{\partial t_2}.
$$
The same observation allows us to rewrite the first two VC ($q=-1$ and $q=0$) in the form that involves among $t$-derivatives also $x$-derivatives (below we use the notation $\tilde{g}(x;\bm t)=\log\tilde{\tau}_k\set{\bm t}$)
\begin{eqnarray}\label{VirTilde-1}
2k^2+2\sum_{m=2}^\infty mt_m\frac{\partial \tilde{g}}{\partial t_m}+2t_1\frac{\partial \tilde{g}}{\partial t_1}+k\nu-2\frac{\partial \tilde{g}}{\partial t_1}+4x\frac{\partial \tilde{g}}{\partial x}=0;\\\label{VirTilde0}
4k\frac{\partial \tilde{g}}{\partial t_1}+2\sum_{m=1}^\infty mt_m\frac{\partial\tilde{g}}{\partial t_{m+1}}-2t_1\frac{\partial\tilde{g}}{\partial x}+\nu\frac{\partial \tilde{g}}{\partial t_1}+2\frac{\partial \tilde{g}}{\partial x}-4x\frac{\partial \tilde{g}}{\partial t_3}=0.
\end{eqnarray}
These two equations give all necessary information. Indeed, from~(\ref{VirTilde-1}) we obtain derivatives over $t_1$:
\begin{eqnarray*}
\left.\frac{\partial \tilde{g}(x;\bm t)}{\partial t_1}\right|_{\bm t=0}	&=&\frac{k(k+\nu)}{2}+2x\frac{\partial\tilde{g}(x;\bm 0)}{\partial x}\;;\\
\left.\frac{\partial^2 \tilde{g}(x;\bm t)}{\partial t_1^2}\right|_{\bm t=0}	&=&\frac{k(2k+\nu)}{2}+2x\frac{\partial\tilde{g}(x;\bm 0)}{\partial x}+4\left(x\frac{\partial}{\partial x}\right)^2 \tilde{g}(x;\bm 0)\;;\\
\left.\frac{\partial^4 \tilde{g}(x;\bm t)}{\partial t_1^4}\right|_{\bm t=0}	&=&3k(2k+\nu)+12x\frac{\partial\tilde{g}(x;\bm 0)}{\partial x}+44\left(x\frac{\partial}{\partial x}\right)^2 \tilde{g}(x;\bm 0)
\\&&\qquad\quad+48\left(x\frac{\partial}{\partial x}\right)^3 \tilde{g}(x;\bm 0) +16\left(x\frac{\partial}{\partial x}\right)^4\tilde{g}(x;\bm 0)\;.
\end{eqnarray*}
Then from~(\ref{VirTilde0}) one can get the mixture derivative over $t_1$ and $t_3$:
$$
\left.\frac{\partial^2\tilde{g}(x;\bm t)}{\partial t_1\partial t_3}\right|_{\bm t=0}=\frac{1}{4x}
\left((4k+\nu)\left.\frac{\partial^2 \tilde{g}(x;\bm t)}{\partial t_1^2}\right|_{\bm t=0}+2\frac{\partial}{\partial x}\left.\frac{\partial \tilde{g}(x;\bm t)}{\partial t_1}\right|_{\bm t=0} -2\frac{\partial \tilde{g}(x;\bm 0)}{\partial x}\right).
$$
Substitution of these results into the KP-equation gives rise to a nonlinear equation in partial derivatives of the function $\tilde{g}(x;\bm 0)$:
\begin{eqnarray}\fl
\qquad-\frac{1}{2}k\left(8 k^2+6k\nu+\nu^2\right)+\frac{3}{2}k(2 k+\nu)\left(2 k^2+k\nu+2\right)x-\nonumber\\\fl\qquad\qquad\qquad\qquad
2\left(1+3 (4 k+\nu) x-6 \left(6 k^2+3 k \nu+10\right) x^2\right)\tilde{g}'\nonumber\\\fl\qquad\qquad\qquad\qquad
-x \left(1+4 (4 k+\nu) x-12 (4k^2+2 k\nu+25) x^2\right) \tilde{g}''\nonumber\\
\fl\qquad\qquad + 216 x^3 \left( \tilde{g}'\right) ^2+288 x^4 \tilde{g}' \tilde{g}''+96 x^5 \left( \tilde{g}''\right) ^2+144 x^4 \tilde{g}^{(3)}+16 x^5 \tilde{g}^{(4)}=0\label{ApD}
\end{eqnarray}

Note that the procedure of joint resolving of KP-equation and VC's fails if we try to apply ``deform-and-study'' approach directly to the integral~(\ref{J_aux}). In this case the $q$-th VC contains the term $\frac{\partial}{\partial t_{q+4}}$, which becomes already at $q=-1$ a derivative over $t_3$. This gap in derivatives makes the system KP-VC unresolvable.
\medskip

{\bf Recurrence relation and some explicit results for $T_{k,0}$.} To derive the recurrence relation for the moments $T_{k,0}$ we substitute, first, the Taylor expansion of the function $\tilde{g}^\nu(x)$ (equipped with an extra index $\nu=\pm 1$):
\begin{eqnarray}\label{g_ansatz}
\tilde{g}_\nu(x)=\log\Xi_k^\nu(x)\;, \quad \Xi_k^\nu(x)=a_k^\nu\sum_{m=0}^\infty(-1)^m \tilde{c}^{\nu}_{k\;m}\frac{x^m}{m!},
\end{eqnarray}
where
$$
 a_k^\nu=\prod_{j=0}^{k-1}\Gamma(1+j)\Gamma\left(1+ j+\frac{\nu}{2}\right),
$$
into equation~(\ref{ApD}).
In the derivation we used a general relation
\begin{eqnarray*}\fl
x^\ell \frac{\partial^s}{\partial x^s}\left(\sum_{j=0}^\infty A_j\frac{(-x)^j}{j!}\right)\cdot \frac{\partial^r}{\partial x^r}\left(\sum_{j=0}^\infty A_j\frac{(-x)^j}{j!}\right)\\=\sum_{j=\ell}^\infty \frac{(-x)^j}{j!}\;
\sum_{m=0}^{j-\ell}\frac{(-1)^{\ell+s+r}j!}{m!(j-m-\ell)!}A_{m+s}A_{j-m+r-\ell}
\end{eqnarray*}
 As the result we obtain the recurrence relation for the coefficients (it is assumed that the summation up to a negative limit is an identical zero) $\tilde{c}^{\nu}_{k\;m}$ ($m\ge1$)
\begin{eqnarray}\nonumber\fl
(2+m)\tilde{c}^\nu_{k\;0}\tilde{c}^\nu_{k\;m+1}=\frac{1}{2} k (2 k+\nu) (4 k+\nu)\sum_{j=0}^m
\left({m}\atop{j}\right)\tilde{c}^\nu_{k\;j}\tilde{c}^\nu_{k\;m-j}
\\\nonumber\fl
+j\sum_{j=0}^{m-1}\left( {m-1}\atop{j}\right)\bigg[\frac{3}{2} k (2 k+\nu) (2k^2+k\nu+2)\tilde{c}^\nu_{k\;j}\tilde{c}^\nu_{k\;m-j-1}+6 (4 k+\nu)\tilde{c}^\nu_{k\;j}\tilde{c}^\nu_{k\;m-j}
\\\nonumber\fl\qquad\qquad\quad\qquad\qquad\qquad\qquad\qquad\qquad\qquad\qquad\qquad\qquad\quad
+\frac{j-1}{j+1}\tilde{c}^\nu_{k\;j+1}\tilde{c}^\nu_{k\;m-j}\bigg]
\\\nonumber\fl
+4j(m-1) \sum_{j=0}^{m-2}\left( {m-2}\atop{j}\right)\bigg[ 3 (6k^2+3k\nu+10)\tilde{c}^\nu_{k\;j}\tilde{c}^\nu_{k\;m-j-1}
-(4 k+\nu)\tilde{c}^\nu_{k\;j+1}\tilde{c}^\nu_{k\;m-j-1}
\\\nonumber\fl\qquad\qquad\quad\qquad\qquad\qquad\qquad\qquad\;\;\;\;
+(4 k+\nu)\tilde{c}^\nu_{k\;j}\tilde{c}^\nu_{k\;m-j}-\frac{1}{4(j+1)}\tilde{c}^\nu_{k\;j+1}\tilde{c}^\nu_{k\;m-j}\bigg]
\\\nonumber\fl
-12j(m-1)(m-2) \sum_{j=0}^{m-3}\left( {m-3}\atop{j}\right) \bigg[(4k^2+2 k\nu+7)\tilde{c}^\nu_{k\;j+1}\tilde{c}^\nu_{k\;m-j-2}
\\\nonumber\fl\qquad\qquad\qquad\qquad\qquad\quad\qquad\qquad\qquad\qquad\qquad\;\;\;\;
-(4k^2+2k\nu+25)\tilde{c}^\nu_{k\;j}\tilde{c}^\nu_{k\;m-j-1}\bigg]
\\\nonumber\fl
-144j(m-1)(m-2)(m-3) \sum_{j=0}^{m-4}\left( {m-4}\atop{j}\right)\bigg[\tilde{c}^\nu_{k\;j+1}\tilde{c}^\nu_{k\;m-j-2}
-\tilde{c}^\nu_{k\;j}\tilde{c}^\nu_{k\;m-j-1}\bigg]
\\\nonumber\fl
+ 16j(m-1)(m-2)(m-3)(m-4) \sum_{j=0}^{m-5}\left( {m-5}\atop{j}\right)\bigg[3\tilde{c}^\nu_{k\;j+2}\tilde{c}^\nu_{k\;m-j-3}
-4\tilde{c}^\nu_{k\;j+1}\tilde{c}^\nu_{k\;m-j-2}
\\\fl\qquad\qquad\qquad\qquad\quad\qquad\qquad\qquad\qquad\qquad\qquad\qquad
+ \tilde{c}^\nu_{k\;j}\tilde{c}^\nu_{k\;m-j-1}\bigg].\label{Tk0_recurrency}
\end{eqnarray}
It can be resolved with the only initial condition $\tilde{c}^\nu_{k\;0}=1$ that follows from the particular choice of the parameter $a_k^\nu$ in the expansion~(\ref{g_ansatz}) and the definition~(\ref{G_def}) of $\Xi_k^\nu(x)$.

The sought moments $T_{\ell,0}$ one can find from~(\ref{G_apear}) by comparison of the the coefficients of Taylor expansions in both sides:
\begin{eqnarray*}\fl
\sum_{\ell=0}^\infty T_{\ell,0}\frac{(-x)^\ell}{\ell!}=\\
\frac{2^{N(N-1)/2}}{\pi^{N/2}\prod_{j=1}^{N}\Gamma(j)}\cases{a_k^+a_k^-
			\sum_{\ell=0}^\infty  \frac{(-x)^\ell}{\ell!}\sum_{m=0}^\ell\left( {\ell}\atop{m}\right) \tilde{c}^+_{k\;m}\tilde{c}^-_{k\;\ell-m},& $N=2k;$\\
			a_k^+a_{k+1}^-\sum_{\ell=0}^\infty  \frac{(-x)^\ell}{\ell!}\sum_{m=0}^\ell\left( {\ell}\atop{m}\right) \tilde{c}^+_{k\;m}\tilde{c}^-_{k+1\;\ell-m},& $N=2k+1$}
\end{eqnarray*}
Then, after a slight massage we arrive to a simple formula
\begin{equation}\label{general_answer}\fl
T_{\ell,0}=\left\langle\big(\tr\bm  z^4\big)^\ell\right\rangle_{GUE_{N\times N}}=
\sum_{m=0}^\ell\left( {\ell}\atop{m}\right) \tilde{c}^+_{k\;m}\tilde{c}^-_{s\;\ell-m}\;,\qquad k=\left\lfloor N\right\rfloor\;,\quad s=\left\lceil N\right\rceil\;,
\end{equation}
Below we reproduce the first three moments
\begin{eqnarray*}
T_{0,0}&=&1;\\
T_{1,0}&=&\frac{N}{4}+\frac{N^3}{2};\\
T_{2,0}&=&\frac{61 N^2}{16}+\frac{5 N^4}{2}+\frac{N^6}{4};\\
T_{3,0}&=&\frac{45 N}{2}+\frac{6517 N^3}{64}+\frac{1101 N^5}{32}+\frac{57 N^7}{16}+\frac{N^9}{8}.
\end{eqnarray*}

\section{Recurrence relations for $T_{k,m}$}
\label{app3}

{\bf $\tau$-function and Virasoro constraints.} To define the $\tau$-function for this case we perform $\bm t$-deformation of the measure in the original integral~(\ref{J2_aux}), so that
\begin{equation}\label{tau_init1}
 \tau_N\set{\bm t}=\frac{1}{N!}\int_{{\cal R}^N}\Delta_N^2(\bm z)
	\prod_{j}^{1\dots N}\exp\left[- z_j^2+y z_j^3-x z_j^4
		+\sum_{k=1}^\infty t_k z_j^k\right]d z_j,
\end{equation}
The extra dependence of $\tau$ on one more additional parameter allows to resolve successfully the KP-equation and the VC. A similar question was discussed in details in the paragraph under equation~(\ref{ApD}).

To derive the VC for the $\tau$-function~(\ref{tau_init1}) we use the transformation of the form
$$
 z_j\to  z_j+\eps z_j^{q+1} \qquad q=-1,0,1\dots,
$$
then the general form of VC reads as follows
$$
\sum_{m=0}^q\frac{\partial{\tau_N}}{\partial t_m\partial t_{m+q}}+\sum_{m=1}^\infty mt_m\frac{\partial{\tau_N}}{\partial t_{q+m}}-2\frac{\partial{\tau_N}}{\partial t_{q+2}}+3y\frac{\partial{\tau_N}}{\partial t_{q+3}}-4x\frac{\partial{\tau_N}}{\partial t_{q+4}}=0.
$$
Observing that
\begin{eqnarray*}
\frac{\partial{\tau_N}}{\partial y}&=&\frac{\partial{\tau_N}}{\partial t_3};\\
\frac{\partial{\tau_N} }{\partial x}&=&-\frac{\partial{\tau_N}}{\partial t_4},
\end{eqnarray*}
we can rewrite the first two VC ($q=-1$ and $q=0$) in the form ($g(x,y;\bm t)=\log{\tau_N}\set{\bm t}$)
\begin{eqnarray}\label{Vir-1}
Nt_1+\sum_{m=2}^\infty mt_m\frac{\partial g}{\partial t_{m-1}}-2\frac{\partial g}{\partial t_1}+3y\frac{\partial g}{\partial t_2}-4x\frac{\partial g}{\partial y}=0;\\\label{Vir0}
N^2+\sum_{m=1}^\infty mt_m\frac{\partial{g}}{\partial t_m}-2\frac{\partial g}{\partial t_2}+3y\frac{\partial g}{\partial y}+4x\frac{\partial g}{\partial x}=0.
\end{eqnarray}
\medskip

{\bf Projection of KP onto the hyperplane $\bm t=\bm 0$.}To perform the projection of the KP-equation~(\ref{KP}) onto the hyperplane $\bm t$ one needs to know the following derivatives
$$
 \left.\frac{\partial^4 g(x,y;\bm t)}{\partial t_1^4}\right|_{\bm t=0},\quad\left.\frac{\partial^2g(x,y;\bm t)}{\partial t_1\partial y}\right|_{\bm t=0},\quad\left.\frac{\partial^2g(x,y;\bm t)}{\partial t_2^2}\right|_{\bm t=0}\quad\mathrm{and}\quad\left.\frac{\partial^2g(x,y;\bm t)}{\partial t_1^2}\right|_{\bm t=0}.
$$
The derivative over $t_2$ can be expressed from~(\ref{Vir0}); substitution of the latter into~(\ref{Vir-1}) helps to find the derivative $\frac{\partial g}{\partial t_1}$; then the necessary projections can be found by subsequent differentiations over $t_1$ and $t_2$. As the result we obtain
\begin{eqnarray*}
\fl \left.\frac{\partial^2g(x,y;\bm t)}{\partial t_1\partial y}\right|_{\bm t=0}&=&\frac{9y}{2} g^{(0,1)}+\left(\frac{9 y^2}{4}-2 x\right) g^{(0,2)}+3 x y g^{(1,1)};
\\
\fl\left.\frac{\partial^2g(x,y;\bm t)}{\partial t_2^2}\right|_{\bm t=0}&=&\frac{15y}{4}  g^{(0,1)}+\frac{9y^2}{4}  g^{(0,2)}+6 x y g^{(1,1)};
\\
\fl\left.\frac{\partial^2g(x,y;\bm t)}{\partial t_1^2}\right|_{\bm t=0}&=&\frac{3y}{16}  \left(63 y^2-88 x\right) g^{(0,1)}+\frac{1}{16} \left(8 x-9 y^2\right)^2 g^{(0,2)}+\frac{3x y}{2}  \left(9 y^2-8 x\right) g^{(1,1)};
\\
\fl\left.\frac{\partial^4 g(x,y;\bm t)}{\partial t_1^4}\right|_{\bm t=0}&=&\frac{45}{256} \left(5696 x^2 y-14544 x y^3+6237 y^5\right) g^{(0,1)}
\end{eqnarray*}\vspace{-20pt}
\begin{eqnarray*}
+\left(\frac{10287 x^2 y^2}{4}-\frac{57429 x y^4}{16}+\frac{331695
y^6}{256}-264 x^3\right) g^{(0,2)} \\+\frac{9}{16} x y \left(3424 x^2-11988 x y^2+6939 y^4\right) g^{(1,1)}\\-\frac{9}{16} x \left(8 x-9 y^2\right)
\left(32 x^2-396 x y^2+297 y^4\right) g^{(1,2)}\\+\frac{9}{128} y \left(8 x-9 y^2\right)^2 \left(63 y^2-88 x\right) g^{(0,3)}-\frac{3}{16} x y \left(8 x-9 y^2\right)^3 g^{(1,3)}\\+\frac{27}{8}
x^2 y^2 \left(8 x-9 y^2\right)^2 g^{(2,2)}+\frac{1}{256} \left(8 x-9 y^2\right)^4 g^{(0,4)}.
\end{eqnarray*}
Above it is assumed that in the right hand side the function $g$ is taken at $\bm t=\bm0$. The standard notation for partial derivatives,
$g^{(k,m)}\equiv\frac{\partial^{k+m}g(x,y)}{\partial^k x\; \partial^m y}$, is also used.

Substituting these terms into KP, eq.~(\ref{KP}), we arrive to a nonlinear equation in partial derivatives on the function $\log \mathcal{J}_n(x,y)$ with the maximal derivative of the forth order and quadratic nonlinear terms. This equation can be rewritten in the form of equation on the function $\mathcal{J}_n(x,y)$ itself. The explicit form of this nonlinear (all terms are quadratic in $\mathcal{J}_n(x,y)$) equation is too cumbersome to be reproduced on paper.
\medskip

{\bf Recurrence relation and some explicit results for $T_{k,m}$.} Substitution of the expansion~(\ref{J_expansion}),
\begin{equation*}
 \mathcal{J}_N(x,y)=\sum_{k,m=0}^\infty(-1)^kT_{k,m}\frac{y^m x^k}{m!k!},
\end{equation*}
gives rise to the sought recurrence relations for the coefficients $T_{k,m}$. They are easier to be handled by using computer rather then a pencil, and here, we do not reproduce them in any form, all necessary calculations were done with the help of computer. 
Below we give results for the first several moments.

\begin{eqnarray*}
T_{0,1}&=&\frac{3 N}{8}+\frac{3 N^3}{2};\\
T_{1,1}&=&\frac{471 N^2}{32}+\frac{225 N^4}{16}+\frac{3 N^6}{4};\\
T_{0,2}&=&\frac{4563 N^2}{64}+\frac{675 N^4}{8}+\frac{27 N^6}{4};\\
T_{2,1}&=&\frac{495 N}{4}+\frac{82335 N^3}{128}+\frac{8673 N^5}{32}+\frac{555 N^7}{32}+\frac{3 N^9}{8};\\
T_{1,2}&=&\frac{25515 N}{32}+\frac{1194939 N^3}{256}+\frac{292383 N^5}{128}+\frac{1323 N^7}{8}+\frac{27 N^9}{8};\\
T_{0,3}&=&\frac{382725 N}{64}+\frac{19566765 N^3}{512}+\frac{2713095 N^5}{128}+\frac{59535 N^7}{32}+\frac{405 N^9}{8}.
\end{eqnarray*}

\end{document}